\documentstyle[12pt,procsla]{article}

\begin{document}

\begin{flushright}
TRI-PP-95-8, March 1995
\end{flushright}

\vskip 1.0cm

\centerline{\normalsize\bf What can we learn from the measurement $R_b \equiv
\Gamma(Z\rightarrow b \overline{b})/\Gamma(Z\rightarrow {\rm hadrons})$?
\footnote{to be appear in the proceedings of Beyond the Standard Model
IV, Granlibakken, December 1994}
}

\vspace*{0.6cm}
\centerline{\footnotesize Daniel Ng}
\baselineskip=13pt
\centerline{\footnotesize\it TRIUMF, 4004 Wesbrook Mall}
\baselineskip=12pt
\centerline{\footnotesize\it Vancouver, B.C. V6T2A3, Canada}
\centerline{\footnotesize E-mail: dng@decu07.triumf.ca}

\vspace*{0.9cm}
\abstract{We examine the effect of new physics on the $R_b \equiv
\Gamma(Z\rightarrow \bar{b}b)/\Gamma(Z\rightarrow {\rm hadrons})$.
Conditions for large contributions are derived.}

\section{Introduction}
The SM is generally in excellent agreement with
experiment, recent results on the left-right asymmetry $A_{LR}$ at SLC
\cite{slcalr} and $R_b=\Gamma(Z\to b\overline{b})/\Gamma(Z\to
\hbox{hadrons})$ measured at LEP \cite{rbref} indicate a possible
disagreement at the 2 to 2.5 $\sigma$ level.

It has been shown in the SM that the $Z-b-\overline{b}$
vertex receives an important contribution from heavy top-quark loops
\cite{akhundov}, leading us to speculate if new
physics may also play a role.
The ratio $R_b$ is a clean test for probing the direct contribution
from new physics.  The reasons are that the ratio is insensitive to QCD
corrections as well as the oblique corrections.
$R_b$ is measured to be $0.2192\pm0.0018$ \cite{rbref},
which disagrees with the theoretical value $0.215$ for $m_t=175$ GeV
predicted by the SM.
In this talk, I would like to concentrate on new effects on $R_b$
due to the new physics.

\section{Direct corrections to $R_b$}
\begin{figure}
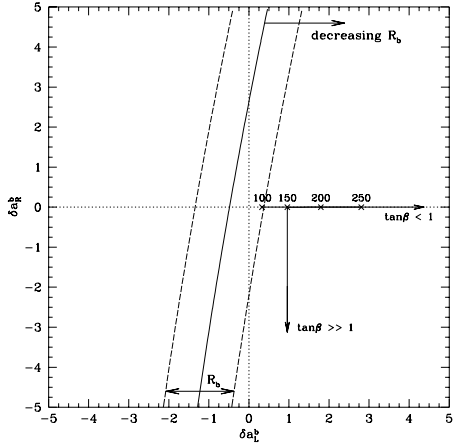

\input{psfig.tex}
\centerline{\psfig{figure=dng.fig,height=2.5in}}
%\vspace{2.5in}
\caption{The $1\sigma$ contour for $R_b$ in the $\delta a_L^b$--$\delta
a_R^b$ plane.  The SM predictions with a heavy top are given by the solid
line with $\delta a_R^b=0$.  Also included in the figure are the small and
large $\tan\beta$ behavior of the vertex corrections in the 2HD model in
the case where $m_t=150$~GeV.}
\label{fig:1}
\end{figure}

With the inclusion of vertex corrections, the $Z-b-\overline{b}$
couplings will be shifted as
$a_{L,R}^f=a_{L,R}^{f,\rm SM}+{\alpha_*\over4\pi s_*^2}\delta a_{L,R}^f$,
where $a_{L,R}^{f,\rm SM}$ are the usual left- and right-handed
couplings and $\delta a_{L,R}^{f,\rm SM}$ are the vertex corrections
which include also the SM top-quark vertex corrections.
Therefore, in the linear order, we obtain \cite{liu/ng}
$R_b\approx 0.2179-0.0021\,\delta a_L^b+0.00038\,\delta a_R^b$,
where $R_b=0.2179$ for $m_t=0$ GeV.  In the SM, $\delta a_R^b=0$,
whereas $\delta a_L^b\sim m_t^2/4M_W^2$ for large $m_t$. In
Fig.~\ref{fig:1}, we show the $1\sigma$ contour of
$R_b$ in the $\delta a_R^b$--$\delta a_L^b$ plane \cite{liu/ng}.
The origin corresponds to the SM with an
unphysical $m_t=0$.  Other values of $m_t$ are indicated on the figure,
showing how the SM expectation value is moved away from the experimental value
of $R_b$.  In particular, there is $\sim2\sigma$ disagreement for
$m_t\approx175$~GeV.  In addition, we see that $R_b$ is more sensitive to
$\delta a_L$ than $\delta a_R$.

\section{New scalars}
Assuming new physics mostly couples to the third generation,
we  consider a single new scalar $\phi$ with arbitrary
isospin($T_3$) and electric charge($Q$) with the interaction
${\cal L}_Y = g\lambda_L\overline{b}_L\phi F_R+{\rm H.c.}$.
The isospins is conserved in the interaction, $T_3(b_L)=T_3(F_R)+T_3(\phi)$.
$F_{L,R}$ may transform differently under the SM to accommodate
both vector and chiral particles.
Thus, we derive
\begin{eqnarray}
\delta a_L^b(\phi)&=&\lambda_L^2[(T_3(b_L)-Q(b)s_*^2)\Theta
-(T_3(\phi)-Q(\phi)s_*^2)(\Theta +\Psi)\nonumber\\
&&\quad+(T_3(F_R)-T_3(F_L))\Delta](M_Z^2;M_\phi^2,m_F^2)\ ,
\label{eq:scalarL}
\end{eqnarray}
where $\Theta$, $\Psi$ and $\Sigma$ are
finite combinations of Passarino-Veltman functions \cite{hvpv} by
\begin{eqnarray}
\Theta(q^2;M^2,m^2)
 &=&q^2[C_{12}+2C_{22}-C_{23}](0,q^2,0;M^2,m^2,m^2)\nonumber\\
\Psi(q^2;M^2,m^2) &=&q^2[2C_{22}-C_{23}](0,q^2,0;m^2,M^2,M^2)\nonumber\\ .
\Delta(q^2;M^2,m^2)&=&m^2C_0(0,q^2,0;M^2,m^2,m^2)\ .
\label{eq:sfunc}
\end{eqnarray}
$\Theta$ and $\Psi$ vanish for $q^2 \to 0$.
This is a consequence of the vector Ward identity.
$\Delta$ is non-vanishing in this limit.
Expanding $\Delta$ in the limit
of $q^2 \ll (m^2,~M^2)$ we obtain.
$\Delta_0(x)= \textstyle{x\over1-x}+\textstyle{x\over(1-x)^2}\log x$,
where $x\equiv m^2/M^2$.  In this limit, $\Theta$ and $\Psi$ are suppressed
by a factor of $q^2/M_\phi^2$.  Therefore, to the lowest order, we
obtain
$\delta a_L^b(\phi)=\lambda_L^2(T_3(F_R)-T_3(F_L))\Delta_0(m_F^2/M_\phi^2)$.
Firstly, this expression vanishes if $F_L$ and $F_R$ carry the same isospin.
This is due to the vector Ward identity.  Therefore, a chiral
fermion in the loop are necessary for a large shift in $\delta
a_L^b$ (and hence $R_b$). Secondly, $\Delta_0$ vanishes
in the limit $m_F\ll M_\Phi$. Therefore, the fermion has to be heavier than
the boson. We also find that
$-1<\Delta_0(x)\le0$ for all $x$.  Thus, the sign of $\delta a_L^b(\phi)$
(or $R_b$) is determined by the isospins
of $F$.  For a scalar $\chi$ with the interaction
${\cal L}_Y = g\lambda_R\overline{b}_R\chi
F_L+{\rm H.c.}$, $\delta a_R^b$ can be obtained from
(\ref{eq:scalarL}) by
$(L\leftrightarrow R)$.

In the two Higgs doublet(2HD) model, $\chi$ is the same scalar as $\phi$ which
is identified as the charged Higgs $H^+$.
$\delta a_{L,R}^b(H^+)$ can be calculated with
$\lambda_L=m_t\cot\beta/\sqrt{2}M_W$ and
$\lambda_R=m_b\tan\beta/\sqrt{2}M_W$.  In Fig.~\ref{fig:1},
we have shown how $\delta a_L^b$ and $\delta a_R^b$
are shifted in the 2HD model relative to the SM with $m_t=150$~GeV.
Note that, the prediction for $R_b$ from the
charged Higgs is always decreased compared to the SM.

\section{New gauge bosons}
Now, we consider the vertex corrections from new gauge bosons with
left- or right-handed coupling, $V_L$-$b_L$-$\overline{F_L}$ or
$V_R$-$b_R$-$\overline{F_R}$, with isospin conservation
$T_3(b_L)=T_3(F_L)+T_3(V_L)$ or $T_3(F_R)+T_3(V_R)=0$ respectively.
For $W=V_L$ and $F=t$, the SM result can be reproduced.
The vertex corrections are calculated to be
\begin{eqnarray}
\delta a_L^b(V_L)&=&[(T_3(b_L)-Q(b)s_*^2){\textstyle{1\over2}}
\Phi+(T_3(V_L)-Q(V_L)s_*^2)[B_0(0;M_{V_L}^2,M_{V_L}^2)-{\textstyle{1\over2}}
(\Phi+\Lambda)]\nonumber\\
&&\quad +(T_3(F_R)-T_3(F_L))\Xi](M_Z^2;M_{V_L}^2,m_F^2)\ ,
\label{eq:vectorL}
\end{eqnarray}
whereas $\delta a_R^b(V_R)$ can be obtained by $(L \leftarrow R)$.
$\Phi$, $\Lambda$ and $\Xi$ are given by
\begin{eqnarray}
{\textstyle{1\over2}}\Phi(q^2;M^2,m^2)&=&
(1+{\textstyle{x\over2}})\Theta(q^2;M^2,m^2)-q^2[C_0+C_{11}]
(0,q^2,0;M^2,m^2,m^2)\nonumber\\
{\textstyle{1\over2}}\Lambda(q^2;M^2,m^2)&=&
(1+{\textstyle{x\over2}})\Psi(q^2;M^2,m^2)+q^2C_{11}
(0,q^2,0;m^2,M^2,M^2)\nonumber\\
&&\qquad\quad-[B_0(q^2;M^2,M^2)-B_0(0;M^2,M^2)]\nonumber\\
\Xi(q^2;M^2,m^2)&=&2m^2C_0(0,q^2,0;m^2,M^2,M^2)
-m^2C_0(0,q^2,0;M^2,m^2,m^2)\nonumber\\
&&\qquad\quad+{\textstyle{x\over2}}[\Theta+\Psi+\Delta](q^2;M^2,m^2)\ ,
\label{eq:wfunc}
\end{eqnarray}
with $x=m^2/M^2$.
We note that the vertex is finite except for the universal piece arising
from $B_0(0;M_V^2,M_V^2)$ which can be removed
by a proper renormalization \cite{kennedy,hollik2}.
Again, $\Phi$ and $\Lambda$
vanish as $q^2 \to 0$, exhibiting the vector Ward identity.
In the limit of $q^2 \ll M_V^2$, we obtain
$\Xi_0(x)=\textstyle{x(-6+x)\over2(1-x)}-
\textstyle{x(2+3x)\over2(1-x)^2}\log x$.
Note that $\Xi_0(x)<0$ for $x \le 0.1$ and $\Xi_0(x)\sim -x/2$ for
large $x$.  Therefore, the vertex corrections to the lowest order is given by
$\delta a_L^b(V_{L,R})=\pm(T_3(F_R)-T_3(F_L))\Xi_0(m_F^2/M_{V_{L,R}}^2)$.
As in the scalar case, in order to have a large vertex correction, $F$
needs to be chiral and heavier than the gauge boson.  The direction
of shift is again determined by the isospins of $F$ at least for a heavy
$F$.  For the SM, we take $F=t$, leading to $\delta a_L^b(W)\sim x/4$.

\section{Conclusion}
In this talk, I have discussed the effects of both new scalars and new gauge
bosons on the vertex corrections in a
model-independent approach.  Two conditions for
large vertex corrections are: fermion in the vertex loop
must be chiral and it must be heavier than the boson in the loop.
We also find that the direction for the corrections is determined by the
isospins of the fermion in the loop.

\vspace{12pt}
I thank Dr. J.T. Liu for collaboration of this work.
This work was supported in part by the Natural Science and Engineering Research
Council of Canada.

\end{document}